\documentclass[twocolumn]{aastex63}

\usepackage{amsmath}
\usepackage{graphicx}
\usepackage{natbib}
\usepackage{mathtools}
\usepackage{amssymb}
\usepackage{xspace}
\usepackage{lineno} 
\usepackage[percent]{overpic}
\usepackage{float}

\newcommand{\degree}{\ensuremath{{}^{\circ}}\xspace}
\newcommand{\pmra}{$\mu_{\alpha{*}}$\xspace}
\newcommand{\pmdec}{$\mu_\delta$\xspace}

\newcommand{\masyr}{mas~yr$^{-1}$\xspace}

\received{\today}
\revised{}
\accepted{}
\submitjournal{ApJ}

\shorttitle{The recent LMC-SMC collision}
\shortauthors{Choi et al.}

\begin{document}

\title{The recent LMC-SMC collision: Timing and impact parameter constraints from comparison of Gaia LMC disk kinematics and N-body simulations}

\author[0000-0003-1680-1884]{Yumi Choi}
\affiliation{Space Telescope Science Institute, 3700 San Martin Drive, Baltimore, MD 21218, USA}
\email{ychoi@stsci.edu}

\author[0000-0002-7134-8296]{Knut A. G. Olsen}
\affiliation{National Optical-Infrared Astronomy Research Laboratory (NOIRLab), 950 North Cherry Avenue, Tucson, AZ 85719, USA}

\author[0000-0003-0715-2173]{Gurtina Besla}
\affiliation{Department of Astronomy, University of Arizona, 933 North Cherry Avenue, Tucson, AZ 85721, USA}

\author[0000-0001-7827-7825]{Roeland P. van der Marel}
\affiliation{Space Telescope Science Institute, 3700 San Martin Drive, Baltimore, MD 21218, USA}
\affiliation{Center for Astrophysical Sciences, Department of Physics \& Astronomy, Johns Hopkins University, Baltimore, MD 21218, USA}

\author[0000-0001-9409-3911]{Paul Zivick}
\affiliation{Mitchell Institute for Fundamental Physics and Astronomy and Department of Physics and Astronomy, Texas A\&M University, College Station, TX 77843, USA}

\author[0000-0002-3204-1742]{Nitya Kallivayalil}
\affiliation{Department of Astronomy, University of Virginia, 530 McCormick Road, Charlottesville, VA 22904, USA}

\author[0000-0002-1793-3689]{David L. Nidever}
\affiliation{Department of Physics, Montana State University, P.O. Box 173840, Bozeman, MT 59717, USA}

\begin{abstract}
We present analysis of the proper-motion (PM) field of the red clump stars in the Large Magellanic Cloud (LMC) disk using the \textit{Gaia} Early Data Release 3 catalog. Using a kinematic model based on old stars with 3D velocity measurements, we construct the residual PM field by subtracting the center-of-mass motion and internal rotation motion components. The residual PM field reveals asymmetric patterns, including larger residual PMs in the southern disk. Comparisons between the observed residual PM field with those of five numerical simulations of an LMC analog that is subject to the tidal fields of the Milky Way and the Small Magellanic Cloud (SMC) show that the present-day LMC is not in dynamical equilibrium. We find that both the observed level of disk heating (PM residual root-mean-square of 0.057$\pm$0.002~\masyr) and kinematic asymmetry are not reproduced by Milky Way tides or if the SMC impact parameter is larger than the size of the LMC disk. This measured level of disk heating provides a novel and important method to validate numerical simulations of the LMC-SMC interaction history. Our results alone put constraints on an impact parameter $\lesssim$10~kpc and impact timing $<$250~Myr. When adopting the impact timing constraint of $\sim$140--160~Myr ago from previous studies, our results suggest that the most recent SMC encounter must have occurred with an impact parameter of $\sim$5~kpc. We also find consistent radial trends in the kinematically- and geometrically-derived disk inclination and line-of-node position angles, indicating a common origin.
\end{abstract}


\section{Introduction}
The Large Magellanic Cloud (LMC) is likely on its first infall towards the Milky Way \citep[e.g.,][]{Kallivayalil2006, Besla2007, vanderMarel2016}, and thus the Milky Way is not likely the main driver shaping the present-day morphology and kinematics of the LMC main body. Consequently, many studies attribute the origin of the asymmetric appearance of the LMC to the close interactions with its nearby companion, the Small Magellanic Cloud (SMC) \citep[e.g.,][]{vanderMarel2001, Yoshizawa2003, Olsen2002, Besla2012, Yozin2014, Besla2013, Besla2016, Pardy2016, Choi2018a, Choi2018b}. In fact, recent studies of the motion of stars and gas in the Magellanic Bridge and the outskirts of the SMC collectively suggest a direct collision between the LMC and SMC about $\sim$100-250~Myr ago \citep[e.g,][]{Zivick2018, Zivick2019, Oey2018, Murray2019, Schmidt2020}. The collision is ``direct" in that the impact parameter of the encounter is expected to be less than the radius of the LMC's stellar disk \citep{Zivick2018}. Thus, constraining the timing and impact parameter of the LMC-SMC collision is the key to constraining the recent dynamical evolution and current morphology of the Magellanic Clouds (MCs). 

Although there have been significant efforts to understand the LMC's internal stellar kinematics, both the lack of a large star sample with accurate and precise 6D phase space information and the lack of baseline dynamical models to compare against the observations have prevented us from developing a complete picture about the LMC's dynamical evolution. 

The advent of large, deep, and MC-targeted photometric surveys over the last decade, e.g., VMC \citep{Cioni2011}, OGLE-IV \citep{Udalski2015}, SMASH \citep{Nidever2017}, has led to an explosion of new discoveries about the MC's star formation histories, dust distribution, and stellar structure, both in the main bodies and the peripheries. The LMC stellar disk is now known to be warped and twisted, such that the inclination and line-of-node position angles vary with galactic radius \citep[e.g.,][]{vanderMarel2001, Subramanian2013, Choi2018a}. Radial variations of the inclination and line-of-node position angles seem to correlate with two significant warps, one at $\sim$2.5~kpc \citep{Olsen2002} and the other at $\sim$5.5~kpc \citep{Choi2018a}, and a tilted off-centered bar \citep[e.g.,][]{Zhao2000, Zaritsky2004, Subramaniam2009, Choi2018a}. 

Many of these and other studies used red clump (RC) stars, which are in the core He-burning stage and have intermediate ages, as tracers \citep[e.g.,][references therein]{Girardi2001, Olsen2002, Subramanian2009, Haschke2011, Girardi2016, Choi2018a, Choi2018b, Gorski2020, Skowron2021}. RC stars are abundant in the LMC out to $\sim$10\degree from its center and trace well the spatial distribution of underlying older stellar populations \citep[e.g.,][]{Choi2018b}. Their uniform stellar core mass at He ignition makes them have very similar effective temperatures and luminosities, which translates to narrow ranges of color and magnitude, respectively.
 
In this paper, we utilize RC stars with proper motion (PM) measurements from the early third data release of \textit{Gaia} \citep{GaiaEDR3}  in order to study their kinematics in the context of the geometrical knowledge already constrained from these stellar populations. To assist our analysis, we rely on a model fit to a sample consisting mainly of red giant branch (RGB) and asymptotic giant branch (AGB) stars, for which we have observed 3D velocities, but no distances as for the RC stars.  We compare the RC PMs against these empirically constrained kinematic models and use hydrodynamic N-body simulations by \citet{Besla2012} to assess the dynamical state of the present-day LMC disk and determine if the data is consistent with theoretical expectations for the disk after a recent encounter with the SMC.

This paper is organized as follows. In Section~\ref{sec:data}, we describe the \textit{Gaia} data for the RC sample used in this study. In Section~\ref{sec:model}, we describe our sample with 3D velocity measurements and kinematic fitting process. We then present the best-fit kinematic model based on the stars that have 3D velocity information. In Section~\ref{sec:results}, we present the internal and residual PM fields of the RC stars in the LMC. We also compare the radial trend of the geometrically- and kinematically measured disk inclination and line-of-node position angles to explore the potential connection between stellar geometry and kinematics. In Section~\ref{sec:comparison}, we describe the numerical simulations used in this study, make a detailed comparison of them with the observed LMC to constrain the recent collision between the LMC and SMC. Section~\ref{sec:conclusion} summarizes our findings and conclusions.   

\begin{figure*}[ht] 
 \centering
      \includegraphics[width=20cm, trim=3.5cm 0cm 0cm 0cm, clip=true]{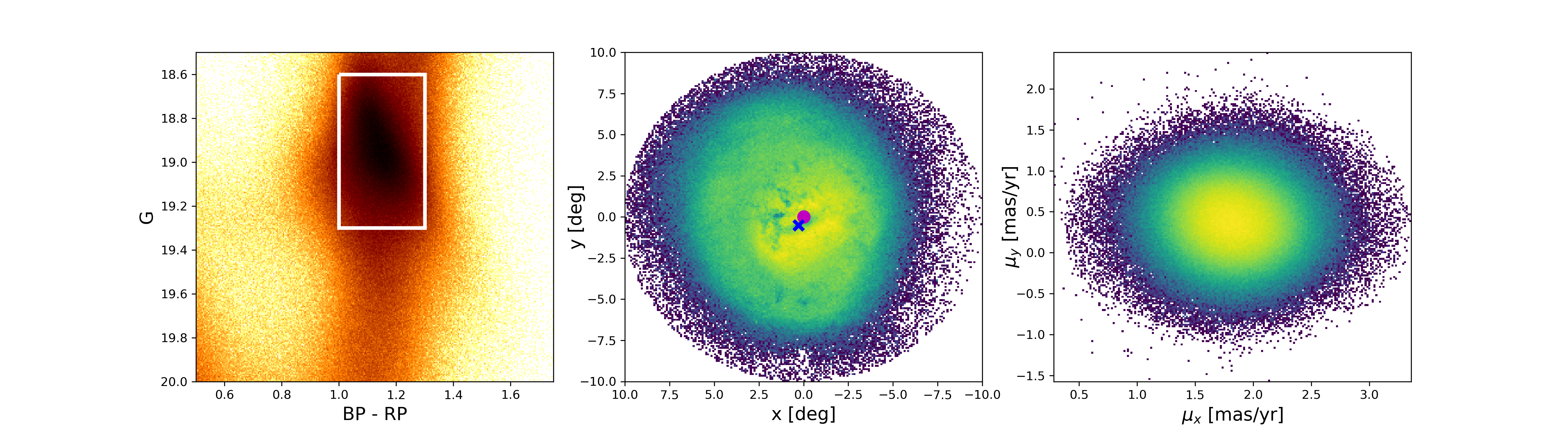}
       \caption{Distributions of the selected RC stars in the Gaia CMD (left), on the sky (middle), and on the PM space (right). The white square in the left panel denotes our CMD selection box. The magenta circle in the middle panel indicates the kinematic center of  ($\alpha_0$,$\delta_0$)=(80.443\degree,-69.272\degree) derived from the observed stellar kinematics in this study (see Section~\ref{sec:model}), while the blue cross indicates the photometric center of ($\alpha_0$,$\delta_0$)=(81.275\degree,-69.783\degree) by \citet{vanderMarel2001b}. 
      \label{fig:rc_selection}}
\end{figure*}

\section{Gaia-selected Red Clump Stars in the LMC} \label{sec:data}
We select the LMC RC stars from the Gaia EDR3 catalog using the following criteria:
\begin{itemize}
    \item 50\degree $\leq \alpha \leq$ 110\degree and  -80\degree $\leq \delta \leq$ -55\degree
    \item $\varpi <$ 0.1 and $\varpi/\sigma_\varpi <$ 5 \citep{Luri2021}
    \item (\pmra - 1.8593)$^2$ + (\pmdec - 0.3747)$^2$ $<$ 1.5$^2$
    \item \texttt{astrometric\_excess\_noise} $<$ 0.2
    \item 1.0 $\leq$ \texttt{phot\_bp\_mean\_mag} - \texttt{phot\_rp\_mean\_mag} $\leq$ 1.3
    \item 18.6 $\leq$ \texttt{phot\_g\_mean\_mag} $\leq$ 19.3
    \item $|C^*| <$ 3$\sigma_{C^*}$ \citep{Riello2021} 
    \item $\rho <$ 10\degree,
\end{itemize}
where $\varpi$ is parallax in mas, \pmra and \pmdec are PMs in Right Ascension (RA; $\alpha$, $\alpha*$=$\alpha$cos$\delta$) and Declination (Dec; $\delta$) in \masyr, respectively, $|C^*|$ is the corrected BP and RP flux excess, and $\rho$ is galactic radius from the LMC kinematic center. Our stringent selection criteria successfully exclude stars with renormalized unit weight error (\texttt{ruwe}) $>$ 1.17 and \texttt{astrometric\_excess\_noise\_sig} $>$ 0.29, assuring that our final catalog consists of single stars with good astrometric solutions \citep[\texttt{ruwe} $<$ 1.4, \texttt{astrometric\_excess\_noise\_sig} $<$ 2;][]{Lindergren2021} and consistent photometry between the G-band magnitude and BP-RP color \citep{Riello2021}. 

Figure~\ref{fig:rc_selection} shows the distributions of the selected RC stars in the Gaia CMD, on the sky, and in the PM space. The above criteria secure a clean LMC RC sample, resulting in a total of 975,637 stars. We do not include significantly reddened RC stars, which are redder and fainter, to keep only brighter RC stars with smaller errors in the PM measurements. The possible contamination of RGB stars \citep[$\sim$10\% level;][]{Choi2018a} does not impact the present study; as we do not expect them to be kinematically distinct from the RC stars. In fact, \citet{Luri2021} revealed almost identical rotation and radial velocity curves for the RC and RGB populations. Following \citet{Luri2021}, we perform our analysis on the (x,y) orthographic projection plane (see their Equations. 1--3). 

Even with significantly improved completeness of \textit{Gaia} \citep{GaiaEDR3,Luri2021}, the RC star census is incomplete along the bar due to crowding; the fainter, the less complete. Incompleteness towards the central regions is not just about the \textit{Gaia} data. All the samples presented in stellar kinematic studies of the LMC in the past and present are incomplete due to crowding. Whether studies are spectroscopic or use PMs, accurate measurements are always available only for the brightest stars from a population, and stars that are blended in crowded areas are always removed from the sample to avoid contamination biases and large uncertainties. However, this incomplete nature of a sample does not matter for stellar kinematic studies because stellar velocity at a given position within a rotating disk is set by the global potential well and the two-body relaxation time in galaxies, which is much longer than the Hubble time. Also, fast and slow moving stars in the same location have identical probabilities to appear blended. Hence, stars in a given population (e.g., RC stars) of different brightness should have statistically identical kinematics particularly when one does not suffer from the small number statistics, which is our case. In other words, for a given stellar population, the mean kinematic properties measured in a sub-region with lower completeness, where inevitably including only brighter RC stars, should be the same as those in the same sub-region that would be measured if completeness were 100\%.

\begin{figure*} 
 \centering
    \begin{overpic}[width=13cm]{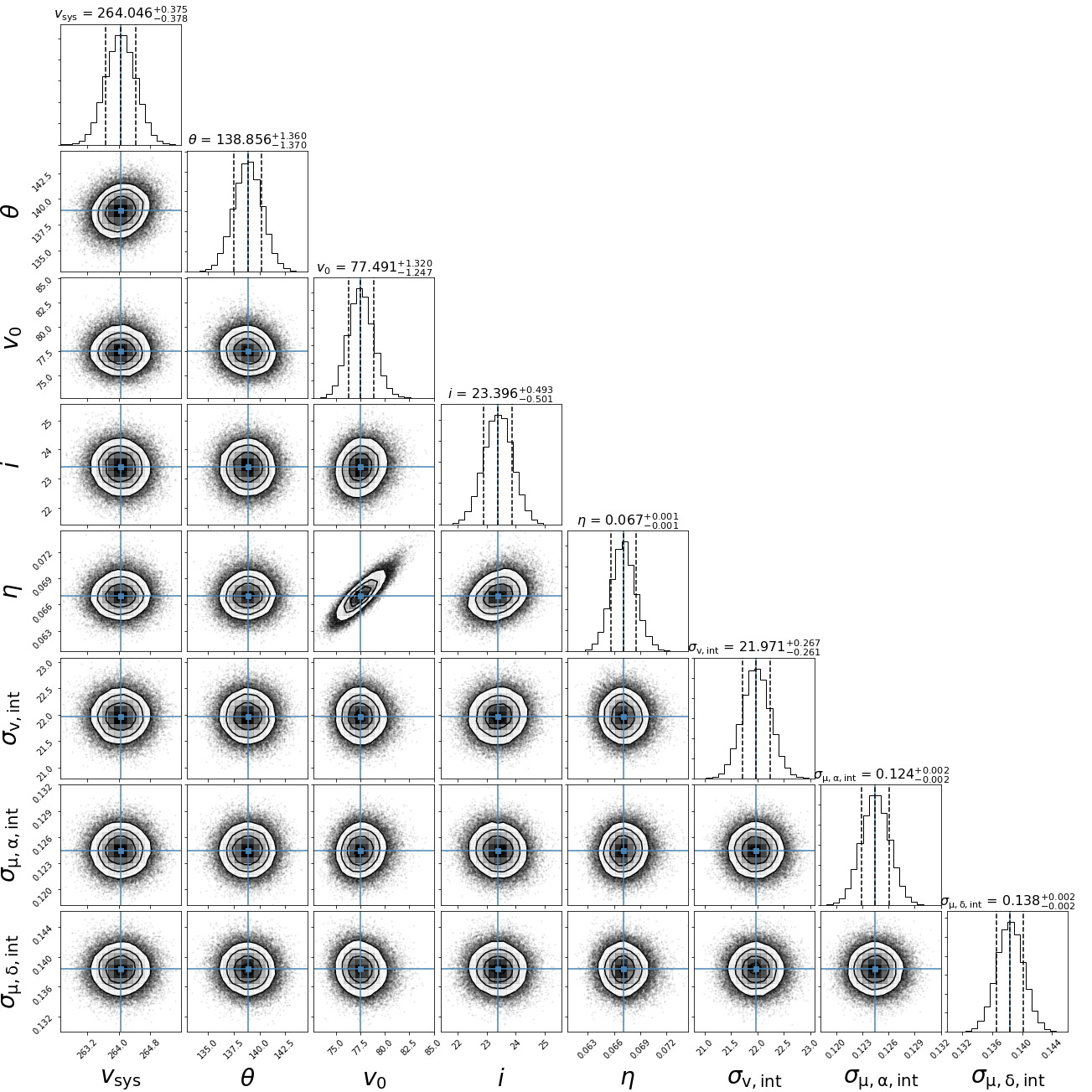}
     \put(60,50){\includegraphics[width=9cm]{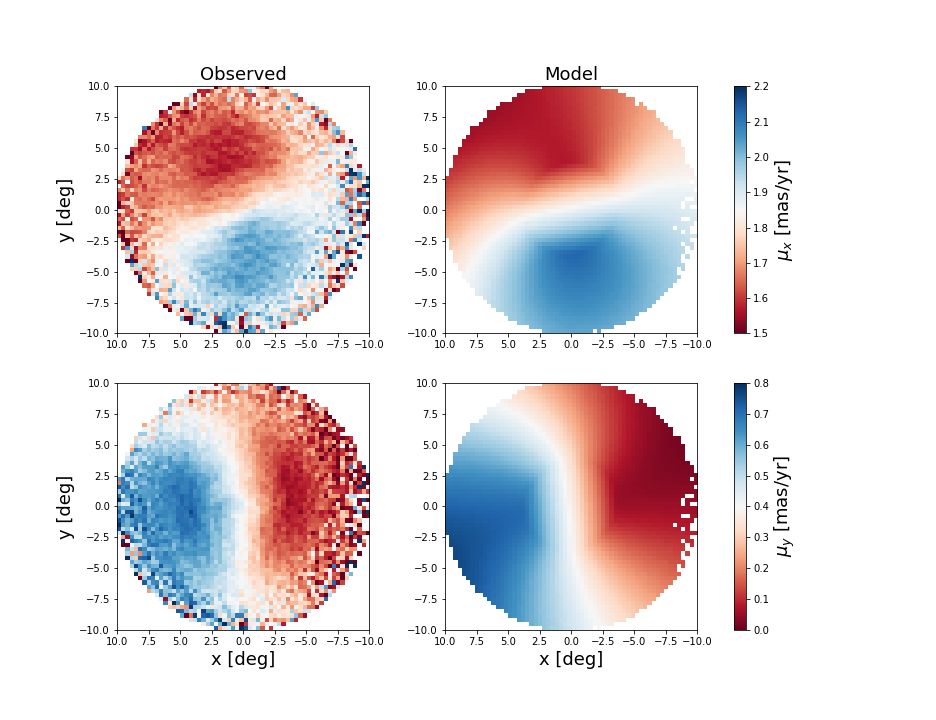}}
    \end{overpic}
       \caption{1D and 2D posterior probability distribution functions of the fitting parameters for our best-fit kinematic model based on RGB and AGB stars that have both PM and line-of-velocity measurements. The inset on the top right corner shows the observed PMs of the RC stars (left panels) and their best-fit model evaluated PMs (right panels) in the x (top rows) and y (bottom rows) directions. The constructed PM field from the best-fit model well describes a disk rotation seen in the observed PM field.
      \label{fig:corner}}
\end{figure*}

\section{Kinematic Model} \label{sec:model}

To investigate the internal kinematics of the LMC's stellar disk, we fit a model to a sample of $\sim$10,000 stars with proper motions from Gaia EDR3 \citep{GaiaEDR3} and line-of-sight velocities from the Hydra-CTIO observations of 4226 stars by \citet{Olsen2011}, 556 unpublished Hydra-CTIO observations (Olsen et al.\ in prep) processed in the identical way to those in \citet{Olsen2011}, and 5386 stars from SDSS DR16/APOGEE-2 \citep{SDSSDR16}. The sample contains predominantly evolved older stars, including RGB and AGB stars, but also $\lesssim$1000 red supergiants.

Our modeling procedure, which is based on the formalism of \citet{vanderMarel2002}, fits up to twelve parameters jointly to the PM and line-of-sight velocity data.  The parameters are the location of the kinematic center in RA and Dec, the bulk transverse motion along the RA and Dec axes, the line-of-sight velocity of the kinematic center, the position angle of the line of nodes, the inclination of the disk, two parameters describing the shape and amplitude of the internal rotation curve, and the velocity dispersion in three orthogonal directions. Throughout, we assume that the LMC disk has no precession or nutation and that the distance to the LMC is 50.1 kpc \citep{Freedman2001}.

To determine the best-fit parameters, we use a combination of the Python package {\tt lmfit} (https://lmfit.github.io/lmfit-py/) to find the maximum likelihood parameter values, and {\tt v3.0.2} of the Markov Chain Monte Carlo package {\tt emcee} \citep{DFM2013,DFM2019} for the final determination of the parameters and their covariances from the posterior probability distributions.  We assume uniform priors for all of the parameters, with conservative limits for each based on physical and empirical expectations. The likelihood function is constructed as the product of Gaussian distributions in each of the observed velocity components (line-of-sight and the two axes of the PM vectors) with respect to the model values.  We use 200 walkers and 50000 steps for the chains, which typically yielded chains 100$\times$ to 200$\times$ longer than the autocorrelation time. Depending on convergence of each fit, the first 600 to 1500 steps are discarded as burn-in.

For this paper, our goal is to derive a model describing the bulk kinematic properties, including internal disk rotation, of our sample of RC stars, the main tracer of this study.  Unfortunately, the LMC RC stars are too faint to obtain spectra for line-of-sight velocity measurements. We thus instead use AGB and RGB stars that have both line-of-sight velocities and proper motions as surrogates for the RC population, as these are known to share bulk kinematic properties with the RC stars \citep{Luri2021}. 

During the course of our fitting, and as has been found in other work \citep[e.g.,][]{Kallivayalil2013, Luri2021}, we found that the results for the LMC's kinematic center and bulk motion depend on the sample from which the fits derive.  In particular, we found that fits to the younger red supergiants yielded a kinematic center that is consistent with the analysis from \citet[see their Table 5]{Luri2021}, allowed for the LMC's \ion{H}{1} gas to trace the rotation curve in a straightforward way, and yielded acceptable fits to the kinematics of the sample of older AGB and RGB stars.  We thus fix the kinematic center and bulk PM to the values obtained for the red supergiant sample, and fit the remaining parameters to the older RGB and AGB stars with radii $<$7\degree.5 from the LMC center.  While a full discussion of the kinematic fits to different subsamples is beyond the scope of this paper, we note that our qualitative results do not depend on the precise center adopted. The full set of our fitting parameters, their priors, and the best-fit for the RGB and AGB sample are summarized in Table~\ref{tab:bestfit}, and the one-dimensional marginalized posterior probability distributions and two-dimensional joint posterior probability distributions for the fitting parameters are presented in Figure~\ref{fig:corner}.

\begin{deluxetable}{lcc}
\tablecaption{The Best-fit Parameters to describe the LMC stellar disk kinematics.
\label{tab:bestfit}}
\tablewidth{0pt}
\tablehead{
\colhead{Parameter} &  \colhead{Prior} &\colhead{Results}}
\startdata
$v_{\rm sys}$ [km~s$^{-1}$] & (250,275) & 264.046$^{+0.375}_{-0.378}$  \\
\hspace{2mm}(systemic velocity) & & \\
$\theta$ [deg]  & (100,190) & 138.856$^{+1.360}_{-1.370}$\\ 
\hspace{2mm}(line-of-node position angle) & & \\
$i$ [deg] & (15,45)  &  23.396$^{+0.493}_{-0.501}$\\ 
\hspace{2mm}(inclination) & & \\
$v_{0}$ [km~s$^{-1}$]  & (20,120) & 77.491$^{+1.320}_{-1.247}$ \\
\hspace{2mm}(in-plane maximum rotation velocity) & & \\
$\eta$  & (0.02,0.2) & 0.067$^{+0.001}_{-0.001}$ \\
\hspace{2mm}(scale radius/distance) & & \\
$\sigma_{\rm v, int}$ [km~s$^{-1}$] & (0,50) & 21.971$^{+0.267}_{-0.261}$ \\
\hspace{2mm}(line-of-sight velocity dispersion) & & \\
$\sigma_{\rm \mu_{\alpha{*}},int}$ [mas yr$^{-1}$] & (0,2) & 0.124$^{+0.002}_{-0.002}$ \\
\hspace{2mm}(proper motion dispersion in $\alpha{*}$) & & \\
$\sigma_{\rm \mu_{\delta},int}$ [mas yr$^{-1}$] & (0,2) & 0.138$^{+0.002}_{-0.002}$ \\
\hspace{2mm}(proper motion dispersion in $\delta$) & & \\
$\alpha_0$ [deg] & Fixed & 80.443 \\
\hspace{2mm}(LMC center in RA) & & \\
$\delta_0$ [deg] & Fixed & -69.272 \\
\hspace{2mm}(LMC center in Dec) & & \\
$\mu_{\alpha{*}},0$ [mas yr$^{-1}$] & Fixed & 1.859 \\
\hspace{2mm}(Center of mass motion in $\alpha{*}$) & & \\
$\mu_{\delta},0$ [mas yr$^{-1}$] & Fixed & 0.375  \\
\hspace{2mm}(Center of mass motion in $\delta$) & & \\
\enddata
\end{deluxetable}

With the best-fit parameters in hand, we evaluate the key parameters used in the present study for our \textit{Gaia}-selected RC stars based on their positions within the disk. The key parameters include the total model PM, the contribution of the LMC's center-of-mass motion to the total PM, and the contribution of internal rotation to the total PM. The inset in Figure~\ref{fig:corner} shows the comparison between the observed PMs and best-fit model-evaluated total PMs of the RC stars in the x and y directions, showing an excellent agreement between the observation and the best-fit kinematic model. This suggests that the observed stellar kinematics of the LMC is well described as an organized circular motion at the first order.

\begin{figure*}[t]
 \centering
      \includegraphics[width=20cm, trim=3.5cm 0cm 0cm 0cm, clip=True]{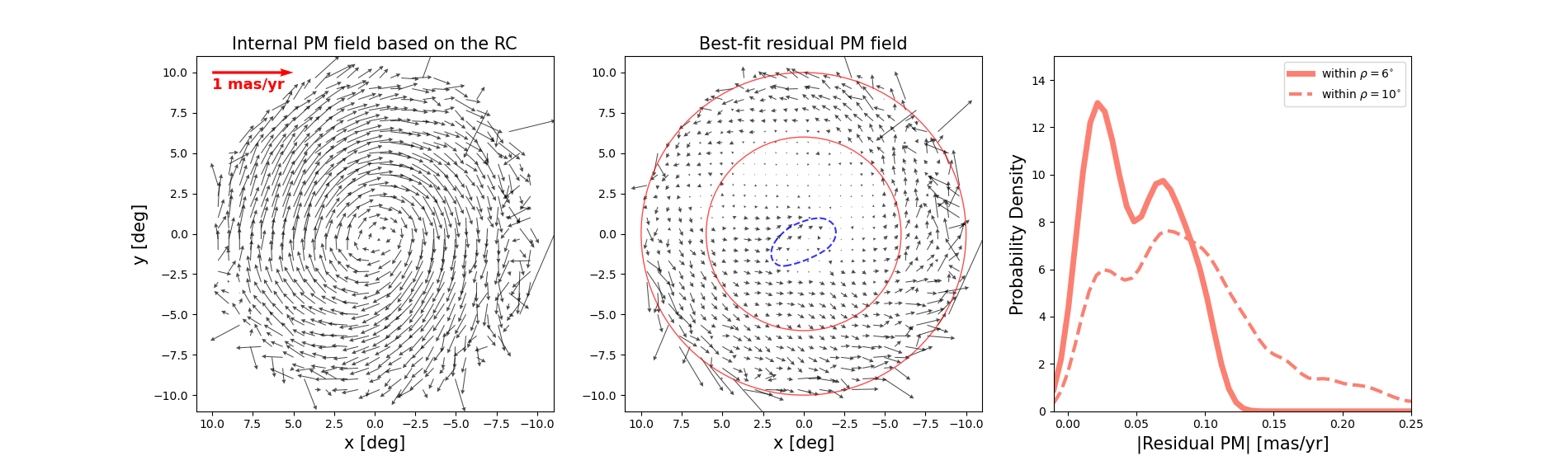}
       \caption{Observed internal motion field of the RC stars (left), residual PM field of the RC stars (middle), and distribution of the residual PM amplitudes of the RC stars within the inner 6\degree(right). We compute the internal motion of the RC stars by subtracting the center-of-mass motion contribution from the observed PM at each position in the LMC disk. The residual PM field of the RC stars is constructed by subtracting the best-fit model PM field, which includes both the center-of-mass motion and the internal rotation motion, from the observed PM field. To guide the eye, we place: a 1~\masyr scale bar in the left panel, galactic radius contours at 6\degree and 10\degree, and a rough outline of the observed bar (blue dashed line). The right panel shows the distributions of the residual PM amplitudes of the RC stars within the inner 6\degree, where we focus our comparison with the simulated galaxies in Section~\ref{sec:comparison}, and the inner 10\degree.   
      \label{fig:obs_pm}}
\end{figure*}

\section{Results} \label{sec:results}
\subsection{Internal and Residual Proper Motions}
Figure~\ref{fig:obs_pm} shows the internal PM field (= observed PM field - the center-of-mass motion field) and the residual PM field (= observed PM field - total best-fit model PM field) in the left and middle panels, respectively. Here the total best-fit model PM field is the combination of the center-of-mass bulk motion and the internal rotation motion fields. On average, the LMC disk exhibits a well-organized clockwise (when seen by the observer) rotation, as shown in the literature. 

Each vector field is constructed by taking mean motion in each $\sim$0.67\degree by $\sim$0.67\degree sub-region, corresponding to $\sim$580~pc at the LMC distance. The size of sub-region is chosen to secure sufficient signal-to-noise ratio (SNR) in the observed residual PM measurements for individual sub-regions; $\sim$95\% of the sub-regions within 6\degree do have SNR $>$ 3 (see below for the justification for the inner 6\degree analysis). The other factor we consider to choose the size of sub-region is the gravitational softening length of 100~pc in the simulations used in this study (see Section~\ref{sec:simulation}). Typically we do not trust much below 3--5 times the softening length because features on smaller physical scales are wiped out by numerical noise. Thus, this optimally chosen sub-region size allows us to measure kinematic features both from the observation and the simulations with high confidence. In addition, rigorous resolution tests show that our qualitative conclusions are robust against the choice of a sub-region size from down to 3$\times$ smaller to up to 4$\times$ larger than the optimal sub-region size. 

Our best-fit kinematic model presented in Section~\ref{sec:model} describes a rotation curve that linearly rises up to the maximum in-plane velocity of 77.491~km~s$^{-1}$ at 3.36~kpc from the center and then remains flat afterward, which is the typically expected pattern for a differentially rotating disk. However, the observed stars actually exhibit a declining rotation curve beyond $\sim$6\degree (corresponding to $\sim$5.25~kpc at the LMC distance) with different slopes for different position angles. The declining behavior of the stellar rotation velocity has been observed in the LMC before \citep[e.g.,][]{Alves2000, vanderMarel2014, Wan2020, Luri2021}, and is partially attributed to elliptical orbits \citep{vanderMarel2001}. 

This position angle-dependent declining rotation curve results in counter-rotating motions with varying magnitude in the residual PM field at larger radius ($>$6\degree) where the observed rotation curve deviates from the best-fit kinematic model (middle panel of Figure~\ref{fig:obs_pm}). Therefore, we limit our analysis to the inner 6\degree in order to correctly evaluate the dynamical status of the LMC disk in a regime where our rotating disk model is valid. 

In the inner 6\degree, the residual PM field shows clear asymmetric features; the residuals are larger in the southern disk than the northern disk, indicating that only the northern inner disk follows organized circular motions. The residual motions near the center might be due to highly non-circular motions around the tilted bar \citep[e.g.,][]{Choi2018a}. In the right panel of Figure~\ref{fig:obs_pm}, we present the Gaussian kernel density estimate (KDE) of amplitude distributions of the residual PMs measured using the RC stars within the inner 6\degree and 10\degree. The kinematic asymmetry is responsible for the bimodality of the distribution of the residual PM amplitudes (right panel of Figure~\ref{fig:obs_pm}). Larger residuals in the outer disk (6\degree$< \rho <$ 10\degree) due to the declining rotation curve significantly contribute to the second peak and broaden the distribution for the inner 10\degree. The RMS of the inner 6\degree distribution is 0.057$\pm$0.002~\masyr (cf. $\sim$0.058~\masyr for that of the RGB stars. See Appendix). The standard error of the residual PMs' mean is $\sim$0.003~\masyr on average, and 94\% of the sub-regions within 6\degree have signal-to-noise ratio greater than 3 in the residual PM measurements. In order to interpret and better understand these observed properties of the residual PM field, we compare the observation against baseline numerical models (see Section~\ref{sec:comparison}).

\begin{figure}[ht]
 \centering
      \includegraphics[width=\linewidth]{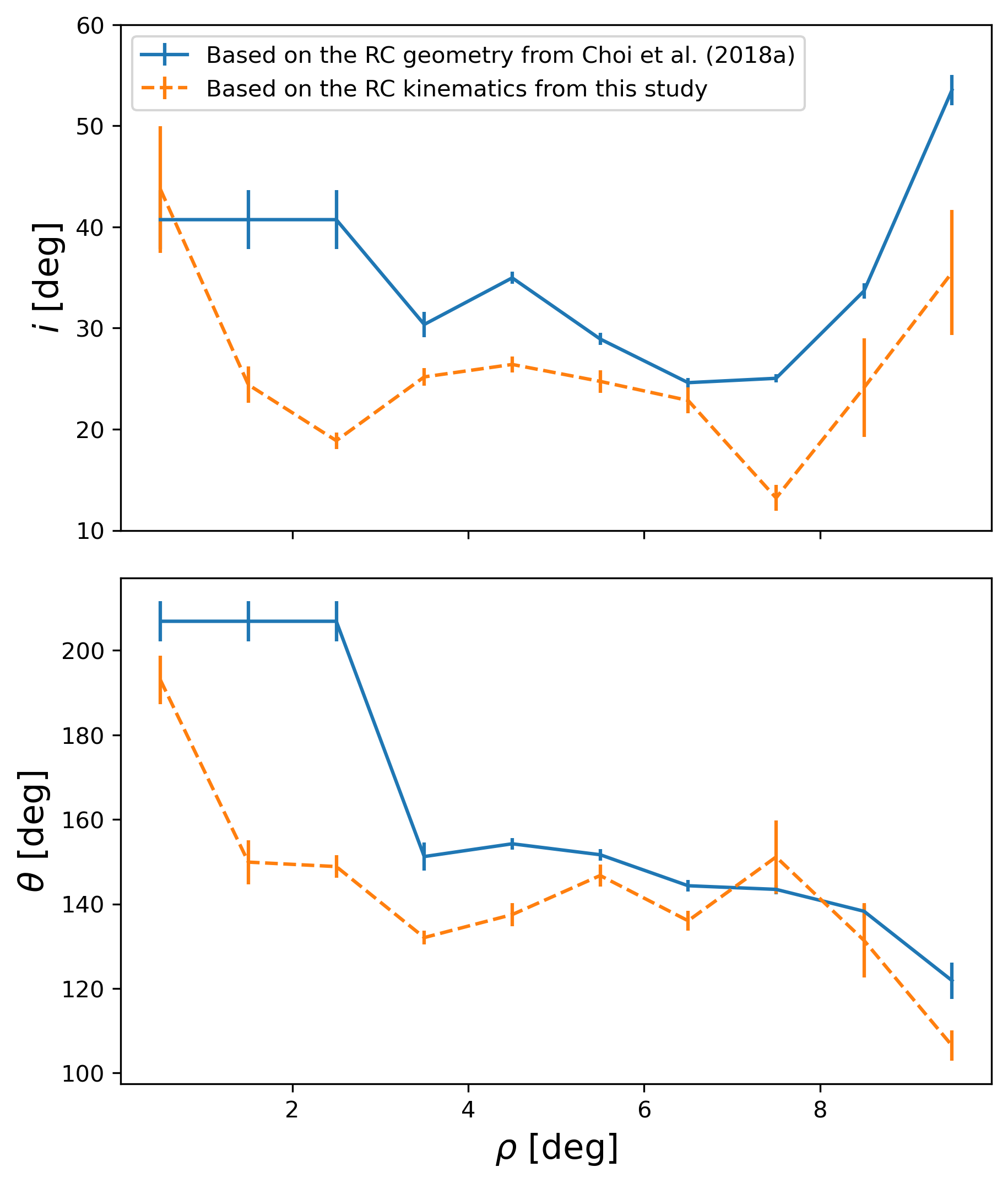}
       \caption{Comparison of the geometrically- and kinematically-measured $i$ and $\theta$ in each 1\degree-width annulus as a function of galactic radius. The geometrically-measured $i$ and $\theta$ values are from \citet{Choi2018a}; they explored the three-dimensional geometry of the RC stars selected from the Survey of the MAgellanic Stellar History data \citep{Nidever2017}. We assumed the measurements in the innermost two radial bins between 0-2\degree, where \citet{Choi2018a} did not make $i$, $\theta$) measurements, are the same as those made in the 2-3\degree radial bin based on the consistency in those measurements between the each annulus and each circle methods as shown in their Figure 13. 
      \label{fig:incl_PA}}
\end{figure}

\subsection{Connection between the kinematic and geometric features in the LMC disk}
Before comparing the observed LMC with simulated LMC models, we first investigate the impact of a warped and twisted disk on our default kinematic modeling, which assumes a single inclination and line-of-node position angles for the entire disk. As shown in the literature \citep[e.g.,][]{vanderMarel2001, Olsen2002, Choi2018a}, the LMC disk is warped and twisted, likely due to the tidal interactions with the SMC. We repeat the kinematic fitting with the same set of RGB and AGB stars described in Section~\ref{sec:model}, but only fit ($i$,$\theta$) in each 1\degree-width annulus, while fixing the rest of the parameters to be the best-fit parameters (Table~\ref{tab:bestfit}). This allows us to see if modeling the LMC disk as a twisted and warped disk could better reproduce the total observed PMs, leading to smaller residual PMs. In each 1\degree-width annulus, we evaluate the model PMs based on the new best-fit ($i$,$\theta$) along with other parameters from the original best-fit and compute new residual PMs. The resulting composite residual PM field, however, still shows asymmetric features with a similar amplitude of the residual PMs. Although these new best-fit ($i$,$\theta$) values for individual annuli are inevitably measured based on a smaller number of stars, no significant changes in the resulting residual PM field indicate that the twists and warps are not the main driver for the LMC's disk to deviate from simple circular motions. 
Nevertheless, this radially-varying ($i$,$\theta$) measurement enables us to explore potential connections between the kinematic and geometric features by making a direct comparison with the ($i$,$\theta$) measurements made purely based on the three-dimensional geometry of the RC stars by \citet{Choi2018a}. We note that this direct comparison is appropriate even though the completeness of the RC sample used in \citet{Choi2018a} and in the current study is different; the Survey of the MAgellanic Stellar History (SMASH) data have much higher RC completeness than the \textit{Gaia} data \citep{Nidever2017}. This is because the stellar kinematics of the RC stars is independent of completeness as discussed in Section~\ref{sec:data}. 

In Figure~\ref{fig:incl_PA}, we compare the kinematically- and geometrically-measured disk $i$ and $\theta$ as a function of galactic radius. The radial trends found in the kinematically-measured $i$ and $\theta$ are broadly consistent with those found in the geometrically-measured $i$ and $\theta$; decreasing $i$ and $\theta$ with galactic radius with a significant turnover in $i$ beyond 7\degree where the southwest stellar warp is dominant. This suggests that the geometric distortions are imprinted in the stellar kinematics (or vice versa depending on which one happens first), and the LMC's disturbed stellar geometry and kinematics are the byproducts of a common event(s). Specifically, \citet{Choi2018a} attributed the radially-varying $i$ and $\theta$ within the LMC disk to the tilted bar and two stellar warps, likely induced by the recent direct collision with the SMC.

\begin{figure*}[ht]
 \centering
      \includegraphics[width=20cm, trim=3cm 2cm 1cm 2cm, clip=True]{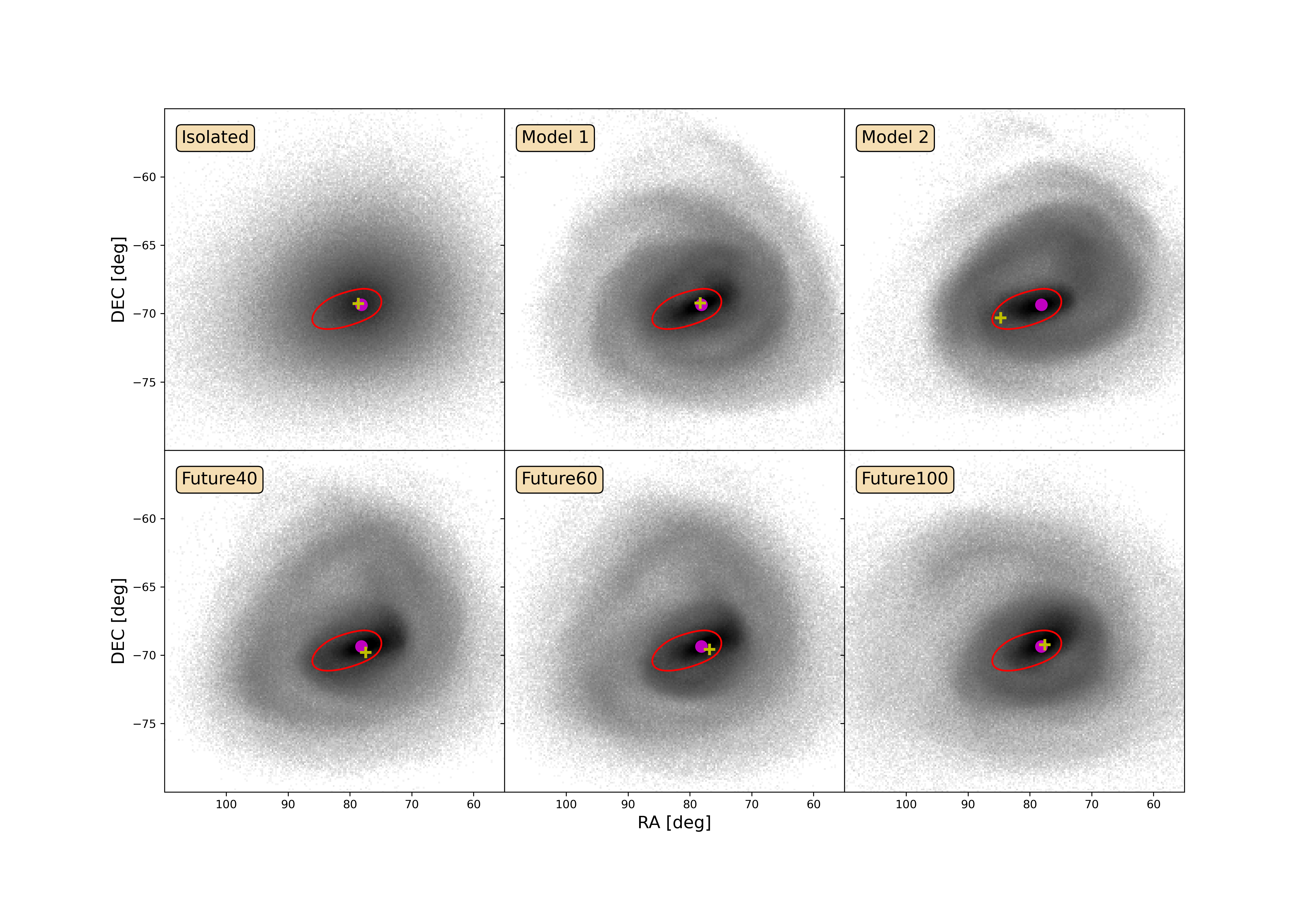}
       \caption{Star count maps using star particles formed $>$1 Gyr ago in the six numerical models considered in this study. In each simulation, the stellar mass per particle is $2500$ M$_\odot$. From top left to bottom right: Isolated LMC, Model 1 (LMC with high impact parameter $>$20 kpc), Model 2 (direct collision $\sim$100 Myr ago), Future40 (direct collision $\sim$140 Myr ago), Future60 (direct collision $\sim$160 Myr ago), and Future100 (direct collision $\sim$200 Myr ago). Each modelled disk is also subject to the tidal field of a Milky Way host (M$_{\rm vir} = 10^{12}$ M$_\odot$) as the LMC enters the Milky Way halo and orbits to its current location over the past 1 Gyr. We rotate each simulated LMC disk to a viewing perspective consistent with the real LMC disk and orientation of the bar. The rough location of the observed bar is outlined by the red ellipse. In each panel we mark the fixed center of mass of the simulated stellar disk with a magenta circle, while the kinematically-derived center (see Section~\ref{sec:model}) is marked with a yellow cross. 
      \label{fig:model_SCmaps}}
\end{figure*}

\begin{figure*}
 \centering
      \includegraphics[width=20cm, trim=3cm 2cm 1cm 2cm, clip=True]{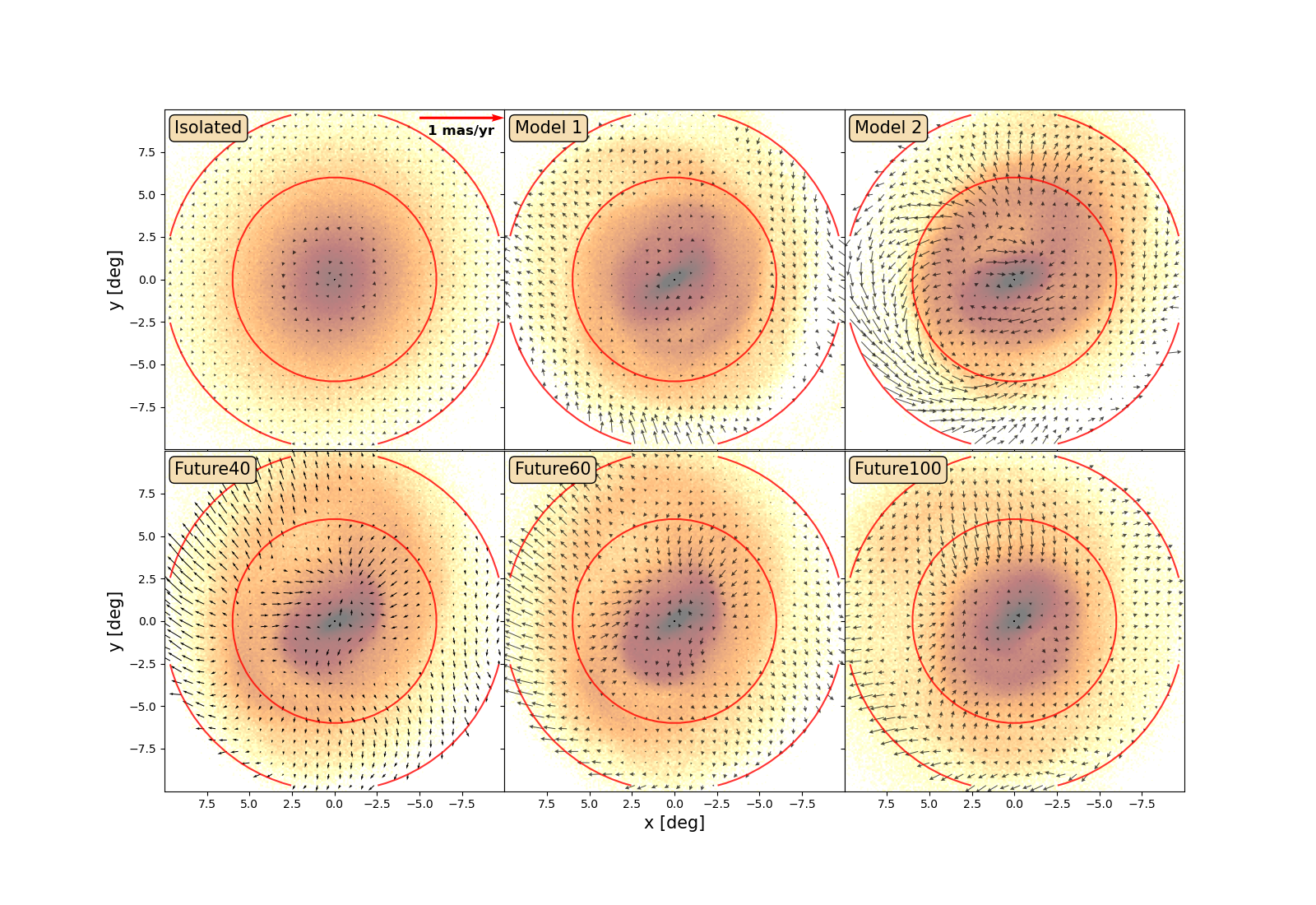}
       \caption{Residual PM fields for the six simulated LMCs: Isolated, Model 1, Model 2, Future40, Future60, and Future100 from top left to bottom right. The red circles denote the galactic radius of 6\degree and 10\degree to guide the eye. The blue dashed line marks the observed bar, as in Figure~\ref{fig:obs_pm}. The residual PM field of the Isolated model (i.e., unperturbed model) shows a low level of residuals in the inner $\sim$3\degree and in the outer $\sim$6\degree. These are the artifacts due to our assumed flat rotation curve in kinematic modeling. Thus, we subtract the Isolated model's residual PM field from the other five models to remove these artifacts.
      \label{fig:model_residualPMs}}
\end{figure*}

\section{Comparison with Theoretical Models} \label{sec:comparison}

\subsection{Simulated LMCs from Besla et al. (2012)}\label{sec:simulation}
\citet{Besla2012} presented numerical hydrodynamic simulations of the interacting LMC and SMC galaxies over the past 6-7 Gyr, including their entry into the Milky Way potential for the first time within the last Gyr. Two models were explored, Model 1 and Model 2, with the primary difference between them being the interaction history of the LMC with the SMC. In Model 1, the SMC remains well-separated from the LMC, with an impact parameter $\geq$ 20~kpc, over their entire interaction history \citep[c.f., the LMC disk size is 18.5~kpc;][]{Nidever2019}, including a duration of 5 Gyr prior to infall into the Milky Way halo. In Model 2, a recent direct collision (impact parameter $\sim$2~kpc) between the Clouds occurs about 100~Myr ago and the Clouds have interacted for a duration of 6 Gyr prior to infall into the Milky Way halo. We note that, in Model 2, the most recent closest approach of the SMC to the LMC (i.e., pericenter) coincides with the time when the SMC crosses the disk plane of the LMC (i.e., disk crossing). Whereas, in Model 1, the disk crossing (pericenter) occurs $\sim$350~Myr (100~Myr) ago when the separation between the MC was $\sim$27~kpc (20~kpc, but 15~kpc below the LMC disk plane).

In each model the LMC's initial dark matter halo is modeled as a Hernquist profile with a total halo mass of $1.8 \times 10^{11}$ M$_\odot$. The simulated stellar mass at the present day is 3.1$\times10^{9}$ M$_\odot$, where the mass per stellar particle is $2500$ M$_\odot$. The SMC is initially modeled as a Hernquist profile with a halo mass of $2.1\times 10^{10}$ M$_\odot$ before it begins interacting with the LMC on an eccentric, decaying orbit. Both models also account for the gravitational perturbations induced by a Milky Way halo mass of $10^{12}$ M$_\odot$ as the Clouds orbit over the past 1 Gyr. For details, please refer to \cite{Besla2012}.

The authors argue that a direct collision is favored as the structure and kinematics of the post-collision LMC and SMC simulations in Model 2 show better consistency with the observed properties of the MC system than in Model 1, both at large and small scales \citep[e.g.,][]{Besla2016, Choi2018a, Choi2018b, Zivick2019}. Given recent PM measurements for the SMC, it is highly improbable that the most recent encounter between the Clouds occurred more than 300~Myr ago or at impact parameters larger than 20~kpc \citep{Zivick2018}. A collision scenario is thus unavoidable, but the exact timing and impact parameter of the encounter remains uncertain.  

In this study we will examine in depth the stellar kinematics of the simulated LMC disk from the Besla 2012 models to gauge whether our study of the observed dynamical state of the real LMC can be used to inform us of the properties (timing, impact parameter) of the most recent LMC-SMC encounter. 

To this end, we extract four additional LMC models from the same simulations by \citet{Besla2012}. One model is for a completely undisturbed LMC disk (``Isolated''), and the other three models are 40~Myr, 60~Myr, and 100~Myr into the future since the Model 2 snapshot (``Future40", ``Future60'', ``Future100''). These models correspond to 140~Myr, 160~Myr and 200~Myr since the direct collision between the MCs. In total, we have six numerical models to compare against our \textit{Gaia} data results: Isolated, Model 1 (impact parameter $\sim$20 kpc, 100~Myr ago), Model 2 (direct collision 100~Myr ago), Future40, Future60, and Future100. Table~\ref{tab:model_summary} summarizes the six simulated LMCs. 

Each of these LMC models have $>$1 million star particles with full 6D position and velocity information in the Galactocentric coordinate system. The simulation also provides rough age estimates for stellar particles formed throughout the simulation. In this study we exclude star particles with ages younger than 1~Gyr old to focus on the intermediate to old stellar populations, facilitating comparison with the observational results based on the RC stars. 

To translate the simulations to the observed frame of reference of the LMC, we first recenter the center of mass and velocity of the simulated LMC stellar particles to be consistent with the \citet{Kallivayalil2013} values, ($X,Y,Z$) = (-1, -41, -28)~kpc and ($V_X$,$V_Y$,$V_Z$) = (-57, -226, 221)~km~s$^{-1}$, and then rotate the model disks to match the observed viewing perspective \citep[$i=$ 25.86\degree, $\theta=$ 149.23\degree;][]{Choi2018a}, corresponding to the normal vector of ($n_{X}, n_{Y}, n_{Z}$) = (0.1332, 0.9628, 0.2348). Finally, we translate the positions and velocities of each star particle in the Galactocentric coordinate system into $\alpha$ and $\delta$ in the ICRS coordinate system, and \pmra, \pmdec, and line-of-sight velocities using the \texttt{astropy.coordinates} package.

The resulting star count maps of the six model galaxies are projected on the sky and presented in Figure~\ref{fig:model_SCmaps}. For Models 1 and 2, we rotate the disks such that the simulated stellar bar is aligned with the observed bar in the LMC when projected on the sky. For Future40, Future60, and Future100, we correct the amount of rotation by accordingly counter-rotating each disk to align the simulated stellar bar to the observed one, assuming a constant rotation velocity of 80~km~s$^{-1}$, which is consistent with the initial conditions for the simulations (``Isolated" LMC). 

In each panel, the red ellipse roughly outlines the location of the observed LMC bar, and the magenta circle denotes the true center of mass of the simulated LMC, while the yellow cross marks the derived center of each simulated LMC from the star particles' kinematics (see Section~\ref{sec:model}). Even in the undisturbed case (Isolated), there is an offset of ($\Delta\alpha$,$\Delta\delta$) = (-0.483\degree, -0.081\degree) between the true center of mass and kinematically derived center. We confirm that this discrepancy originates from random sampling of star particles with number matched to our observed samples, hinting at the difficulty of defining the LMC center accurately with observational data. The root-mean-square (RMS) values of offsets between the true center of mass of the simulated LMC and the kinematically derived center of each model are 2.765\degree and 0.759\degree in $\alpha$ and 0.448\degree and 0.234\degree in $\delta$ when including and excluding the most strongly perturbed LMC disk (Model 2), respectively. This clearly indicates that the strong direct collision with the SMC can significantly displace the kinematic center from the center of mass for a relatively short period of time since the impact ($\sim$100~Myr). The RMS values of the offsets calculated including Model 2 are significantly larger compared to the range of reported kinematic centers for the observed LMC \citep[$\sigma_{\alpha}$=1.259\degree, $\sigma_{\delta}$=0.359\degree; e.g.,][]{vanderMarel2014, Luri2021}. This analysis indicates that Model 2 (impact parameter $<$ 2 kpc within 100 Myr) results in an LMC disk with a kinematic center that strongly disagrees with observations. The results for Future40, Future60 and Future100 indicate that within an additional 40-100 Myr Model 2 does have sufficient time to settle to a dynamical state more consistent with observations. As such, a direct collision scenario is not ruled out, despite the large offsets exhibited in Model 2.

\begin{deluxetable}{lcc}
\tablecaption{Summary of the six simulated LMCs.
\label{tab:model_summary}}
\tablewidth{0pt}
\tablehead{
\colhead{Model Name} &  \colhead{Impact parameter} &\colhead{Impact Timing} \\
& [kpc] & [Myr ago]}
\startdata
Isolated & None & None \\
Model 1 & 20 & 100 \\
Model 2 & 2 & 100 \\
Future40 & 2 & 140 \\
Future60 & 2 & 160 \\
Future100 & 2 & 200 \\
\enddata
\end{deluxetable}

\subsection{Residual Proper Motions of Simulated LMCs}\label{sec:residuPM_models}
For the six simulated galaxies that are translated to the observed frame (Section~\ref{sec:simulation}), we also conduct the kinematic modeling procedure to stellar particles that have six-dimensional phase space information and derive the best-fit kinematic model for each simulated galaxy. We then evaluate the best-fit model predicted PM components for individual star particles. 

Figure~\ref{fig:model_residualPMs} shows the residual PM fields for the six model galaxies. As described earlier, we compute observables (PMs on sky and line-of-sight velocity) for individual star particles in each simulated LMC. Using these simulated observables, we perform kinematic modeling for the six simulated LMCs and construct the residual PM fields based on the best-fit kinematic model for each simulated LMC. We fit all twelve parameters for randomly selected $\sim$10,000 star particles in each of the simulated LMCs. 

The details of the PM residual fields for the simulated LMCs in Figure~\ref{fig:model_residualPMs} depend on many details of the LMC-SMC interaction histories that are yet poorly constrained, e.g., the masses of the Milky Way, LMC, and SMC, the pre-collision structure of the galaxies, the pre-collision relative spin orientation of the galaxies, etc. Given the limited set of models assessed in this paper, it is therefore not surprising that the details of the observed residual field are not fully reproduced by any of the simulated LMCs. Instead, the amplitudes of the residuals are a direct measure of the disk heating induced by tidal interaction. This is less dependent on the details of the individual galaxies, and is more directly driven by the impact parameter (smaller value yields more heating) and time that has passed since the collision, which enables post-collision cool-down. Since the impact parameter and timing are the orbital characteristics that we are most interested in here, we focus on a quantitative comparison of the amplitude distribution of the residuals, and use the residual field itself only as an additional qualitative constraint (e.g., symmetric vs. asymmetric). 

The Isolated model shows tiny residual PMs (rms of 0.022~\masyr) particularly in the central 3\degree and beyond 6\degree (top left in Figure~\ref{fig:model_residualPMs}). These features turn out to be the artifacts arising from a difference between the parameterized rotation curve assumed in the kinematic modeling and the resulting rotation curve from the mass profile for the initial, unperturbed simulated LMC disk described in Section~\ref{sec:simulation} \citep[see also][]{Besla2012}. We, therefore, correct the residual PM fields of the other five simulated galaxies for the artifacts by subtracting the Isolated model's residual PM field and present the corrected fields in Figure~\ref{fig:model_residualPMs}. After making this correction, the Isolated is no longer utilized. 

To be consistent with the observational data, we limit our analysis of the five simulated galaxies to the inner 6\degree as well. This also minimizes the potential bias in the residual PM fields due to uncertainties in the importance of the Milky Way tidal field (i.e., the uncertainties in the mass of the Galaxy), which will largely impact the outskirts of the LMC disk. In addition, the future models (Future40, Future60, and Future100) interact with the Milky Way for a longer period of time than Models 1 and 2, leading to inconsistent Milky Way tidal fields in the outer disk. 

Non-uniform residual PM features in the inner 6\degree are found in the all five simulated LMCs, which is consistent with the observational data (see Figure~\ref{fig:obs_pm}). However, the strength of the residuals vary across the models. Model 1 (LMC-SMC impact parameter $\sim$20~kpc about 100~Myr ago) is left with insignificant residuals, making it consistent with a marginally perturbed disk. On the other hand, the other simulated galaxies maintain more prominent residuals and asymmetric kinematic features in the inner 6\degree. Specifically, Model 2 (direct collision with an impact parameter of 2 kpc about 100 Myr ago) shows net streaming motions from West to East and from East to West under and above the bar, respectively, and then diverging motion to the North and East. The Future40, Future60, and Future100 models, roughly speaking, present a net inward motion toward the central disk in the inner 3--6\degree, which is not clear in the data (see Figure~\ref{fig:obs_pm}).

\begin{figure}
 \centering
      \includegraphics[width=\linewidth, trim=0cm 1.5cm 0cm 2.5cm, clip=True]{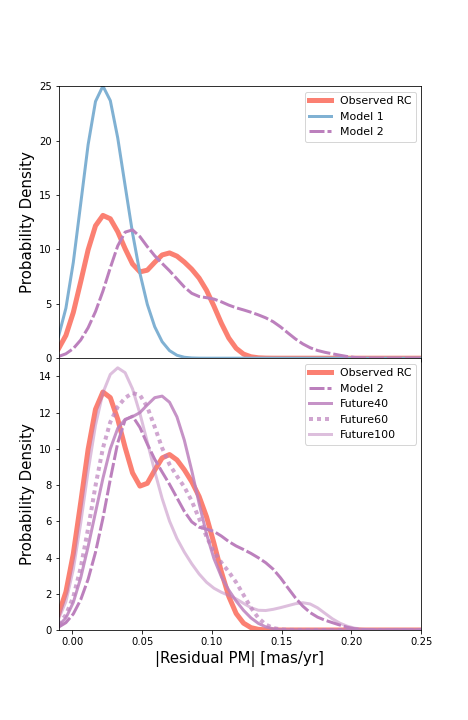}
       \caption{Gaussian KDEs showing distributions of the residual PM amplitudes within inner 6\degree for the observed RC and RGB stars along with those for the five simulated LMC galaxies that are calibrated using the Isolated model residual PM field.
      \label{fig:1D_residualPMs}}
\end{figure}

In Figure~\ref{fig:1D_residualPMs}, we compare the KDEs of the residual PM amplitudes for the observed RC stars and the five simulated LMCs within the inner 6\degree where there are no complexities associated with rotation curve modeling, making our comparison straightforward. The top panel compares the observation with the results from Model 1 and 2, while the bottom panel compares with the Model 2 and its future snapshots (Future40, Future60, and Future100). The RMS values of the each model's residual PM distributions are 0.027~\masyr, 0.084~\masyr, 0.062~\masyr, 0.062~\masyr, 0.067~\masyr for Model 1, Model2, Future40, Future60, and Future100, respectively. 

The distribution of the residual PM amplitudes of Model 1 is completely inconsistent with the observation in terms of its morphology and RMS. Model 1 seems to be perturbed much less compared to the observed LMC disk; it has significantly smaller RMS and no evidence for kinematic asymmetry (i.e., well described as a single Gaussian). On the other hand, Model 2 shows a clear evidence of multiple peaks, a sign of kinematic asymmetry, as seen in the observed data. Nevertheless, Model 2 is dynamically too hot (significantly larger RMS) to explain the observed LMC's disk, suggesting that more time is required for Model 2 to dynamically cool down to be consistent with the observation. Thus, the dynamical status of the present-day LMC likely lies between Model 1 and 2. 

In the bottom panel of Figure~\ref{fig:1D_residualPMs}, we compare the observations and the results from Model 2 (100~Myr after the collision) and its future snapshots, Future40 (140~Myr after the collision), Future60 (160~Myr after the collision), and Future100 (200~Myr after the collision), to more closely investigate the time evolution of the LMC stellar kinematics after a direct collision with the SMC. The residual PM amplitudes rapidly decrease from Model 2 to its future snapshots. The perturbation in Model 2 is significantly damped by $\sim$25\% during the first 40~Myr and then remains more or less the same for the next 60~Myr with a slight increase in Future100. In terms of RMS values, all future models are consistent with each other and more perturbed than Model 1. In addition, they show a hint of kinematic asymmetry via a few bumps or broadening in their distributions, which is consistent with the observation. However, contrary to Future40 and Future60 as well as the observation, Future100 show an significantly extended tail towards higher residuals as much as Model 2, but with significantly lower probability density, which makes its RMS value increase compared to Future40 and Future60 despite its narrower distribution in the smaller residual regime. On average, the dynamical status of the three future models is consistent with that observed, suggesting that the present-day LMC should have evolved for at least 140~Myr since the last collision if the collision was direct with an impact parameter of 2~kpc. Utilizing the smallest probable impact parameter from the LMC-SMC analytic orbit modeling of $\sim$2~kpc \citep{Zivick2018}, this required time of 140~Myr sets the lower limit to the timing of such a collision.

\subsection{Constraining the recent collision with the SMC}\label{sec:constraining}
A clear trend from the residual PM amplitudes of the five simulated LMCs is that the degree of perturbation decreases with either longer impact timing $T$ (i.e., longer elapsed time since the last encounter between the MCs) or larger impact parameter $P$ (i.e., larger separation between the LMC and SMC at the time of impact). For example, an LMC disk with a direct collision with a more modest impact parameter (e.g., 10~kpc) than Model 2 would have had less disk heating initially, and hence would have needed less time to cool down to reproduce the observed amplitude residuals than Model 2. To analytically describe this trend, we fit a simple planar model to log of the RMS ([\masyr]) of the residual PM amplitude distribution as a function of impact timing in Myr and impact parameter in kpc, which yields the correct asymptotic behavior towards longer impact timing and larger impact parameter. The best-fit plane based on the five simulated LMCs is 
\begin{equation}
    log_{10}(\rm RMS) = -0.000946\,T - 0.0274\,P - 0.974.
\end{equation}

\begin{figure}
 \centering
      \includegraphics[trim=0cm 2cm 0cm 2cm, clip=True, width=\linewidth]{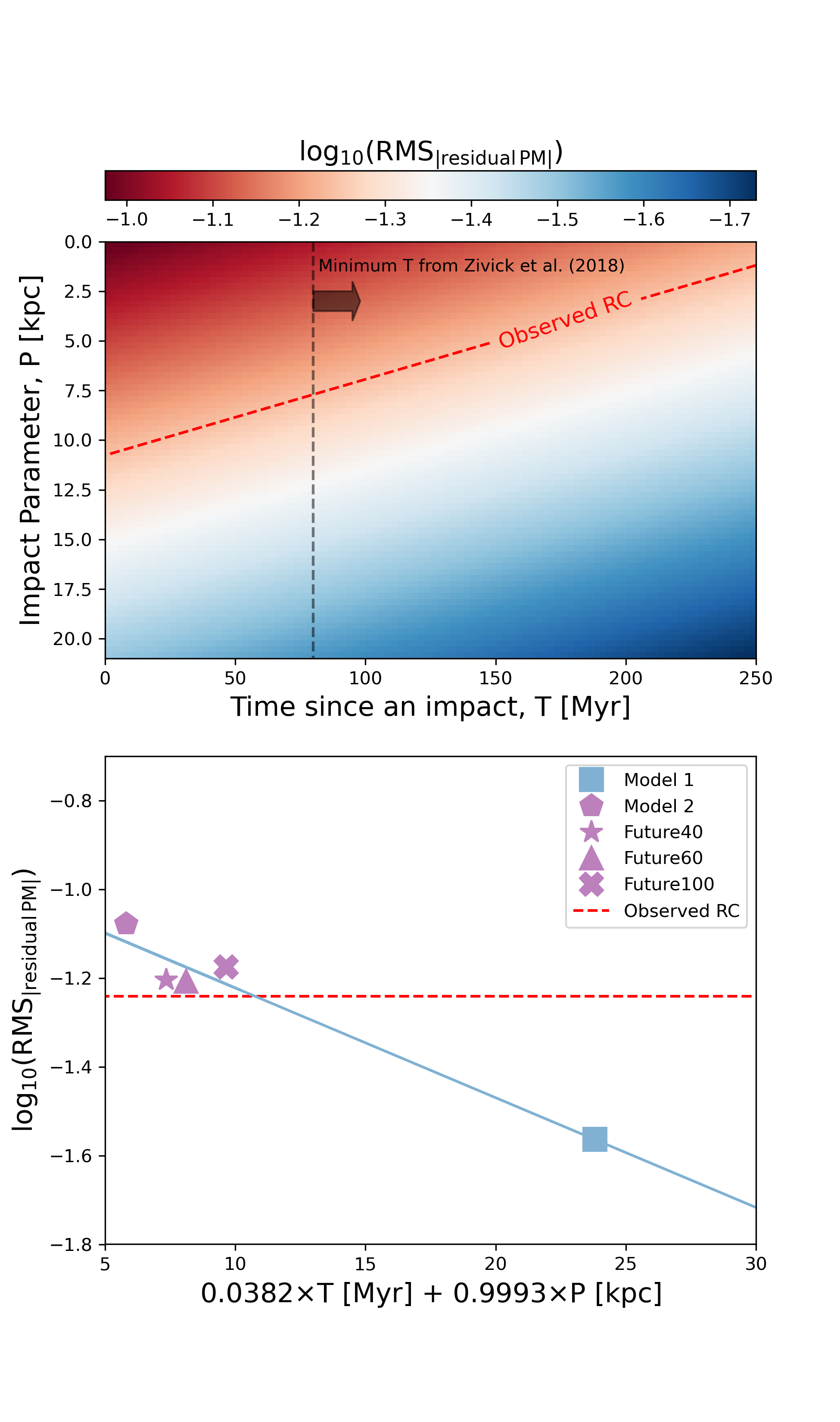}
       \caption{\textit{Top:} Predicted log10 of a RMS of a residual PM amplitude distribution as a function of impact timing (T) and impact parameter (P). The prediction is made from the best-fit plane of (T, P, log10(RMS)) using the five simulated LMCs. The red dashed line indicates the 1-dimensional family of (T, P) that is consistent with the RC-based observational results. The black dashed line denotes the minimum T from the LMC-SMC orbit modeling \citep{Zivick2018}. \textit{Bottom:} The edge-on view of the best-fit plane (blue solid line) and distribution of the five simulated LMCs around the plane. The observation agrees reasonably well with the all three future models (Future40, Future60, and Future100).
      \label{fig:paramspace_fit}}
\end{figure}

The top panel of Figure~\ref{fig:paramspace_fit} shows the predicted RMS values in log scale as a function of (T, P). We depict the possible 1-dimensional family of (T, P) that is consistent with the observations. With the caveat that this plane fitting is based on very limited parameter space and thus should be considered as the first order approximation, we find that the observations disfavor a distant encounter scenario with an impact parameter $>$20~kpc, but favor a close impact scenario with an impact parameter $<$10~kpc. Furthermore, we find that perturbations induced by an SMC collision $\sim$250 Myr ago (with any impact parameter larger than 2~kpc) will always be weaker than that observed. This is consistent with the upper limit placed on the impact timing and impact parameter for the LMC-SMC collision inferred from orbit modelling using the observed PMs of the Clouds \citep{Zivick2018}. We also find that, for the no direct collision case like Model 1 (i.e., impact parameter $>$ LMC disk), the observed level of disk heating cannot be reproduced even if we observed the LMC right after the encounter, which is inconsistent with the observed relative separation ($\sim$20~kpc) and velocity ($\sim$100~km~s$^{-1}$) between the LMC and SMC \citep{Zivick2018}. The minimum probable T from orbit modeling is $\sim$80~Myr \citep[][their figure 12]{Zivick2018}. Combining this with our results places an upper limit on the impact parameter of $\sim$7.5~kpc.  

The edge-on view of the best-fit plane is also presented in the bottom panel of Figure~\ref{fig:paramspace_fit} along with the positions of the five simulated LMCs and the observation. If we take this simple best-fit plane as face value, the observation best agrees with Future100. However, given our small sample of simulated LMCs with very limited coverage of the parameter space, there is insufficient evidence to exclude Future40 and Future60. Furthermore, all the Future models reside close to each other in the (T, P, log10(RMS)) space. Although we can rule out Models 1 and 2.  
This analysis places a lower limit to the timing of the collision, given an impact parameter. In particular, if the collision was direct ($<$5 kpc; Model 2, Future40, Future60, Future100), then the collision cannot have occurred within $\leq$150~Myr. Based on LMC-SMC orbit modeling, the most probable impact timing is 147$\pm$33~Myr ago \citep[][their figure 12]{Zivick2018}. From our plane fitting, this would suggest that the collision would have occurred with the impact parameter of $5_{-1}^{+1.5}$~kpc. This impact parameter, inferred based on the PM residuals, is also consistent with the expected impact position of the collision from orbit modelling (7.5$\pm$2.5~kpc) within uncertainty. 

While Model 1 and 2 are isolated in this space, the observation is clustered around Future40, Future60, and Future100. As expected from its KDE presented in Figure~\ref{fig:1D_residualPMs}, no significant asymmetric feature in the residual PM field is found in Model 1, indicating that a direct collision is necessary to develop the asymmetric features in the stellar kinematics measured in the observational data. Together with the RMS values, the asymmetry further supports that the dynamical state of the Future models are similar and consistent with the observed LMC stellar disk. In conclusion, the present-day dynamical status of the LMC disk requires that the most recent encounter with the SMC be direct, with an impact parameter $\leq$ $\sim$10~kpc and impact timing of $<$250~Myr, based on our analysis \textit{alone}. If we adopt the impact timing constraint of $\sim$140--160~Myr ago from analytic orbit modeling of the LMC-SMC \citep{Zivick2018} and from the star formation history in the Bridge \citep[e.g.,][]{Harris2007, Noel2015}, then the impact parameter must be $\sim$5~kpc. This necessarily means that the recent close encounter also coincides with the time when the SMC crosses the LMC disk plane.

Note that the orbit modeling presented in studies like \cite{Zivick2018} are analytic calculations that do not account for the distortions to the dark matter halos of the Clouds during their interaction; this will influence the true range of plausible impact parameters and timing of the collision. As we have illustrated through comparison to the \cite{Besla2012} models, the PM residuals presented in this paper provide a novel way to validate the assumed impact parameter and timing of the LMC-SMC collision when combined with numerical modeling.

\section{Summary and Conclusion} \label{sec:conclusion}
We explore the kinematics of the RC stars of the LMC selected from the \textit{Gaia} EDR3 catalog. Using the kinematic model best describing the three-dimensional motion of old stellar populations, we construct the residual PM field for the RC stars by subtracting the center-of-mass motion field and the internal rotation motion field from the observed PM field. We focus on the residual PM field in the inner 6\degree where our rotation curve modeling is valid and there are minimal perturbations from Milky Way tides. We also derive the disk inclination and the line-of-node position angles as a function of galactic radius based on stellar kinematics in each 1\degree-width annulus, and compare the results to those based on the three-dimensional geometry of the RC stars. Our main findings from the kinematic analysis of the RC stars in the LMC are: 
\begin{enumerate}
    \item The resulting residual PM field reveals asymmetric patterns, including: a complex residual PM field around the bar, larger residual PMs in the southern disk than the northern disk, and counter-rotating residual motion with varying amplitude as a function of position angle. 
    \item The RMS of the residual PM amplitude distribution, a proxy for disk heating, is 0.057$\pm$0.002~\masyr. 
    \item The radial trends of kinematically- and geometrically-derived inclination and line-of-node position angles are consistent to each other, indicating that the perturbed kinematics and distorted geometry likely have the same origin. In particular, a rapid increase in the geometrically-derived inclination beyond $\sim$7\degree is attributed to the outer stellar warp in the southwest part of the LMC disk, a likely byproduct of the recent direct collision with the SMC \citep{Choi2018a}. The same behavior in the kinematically-derived inclination suggests that the tidal event that induced the stellar warp has also been imprinted in the stellar kinematics.
\end{enumerate}
 
To assess the dynamical state of the present-day LMC disk, we compare the observed stellar kinematics to five numerical simulations of an LMC with different interaction histories with the SMC in a first infall scenario \citep{Besla2012}. Following the same methodology applied to the observational data, we obtain the residual PM fields and measure the RMS of the residual amplitude distribution for each five simulated LMCs. We find that the level of disk heating and asymmetric feature of the residual PM field of the present-day LMC disk are consistent with those exhibited by the Future models (Future40, Future60, Future100) in which an LMC underwent a direct collision with the SMC with an impact parameter of 2~kpc 140-200~Myr ago. Although asymmetric feature exists in Model 2 (only 100~Myr after the direct collision with the SMC), its disk is dynamically too hot (RMS of 0.084~\masyr) to explain the observational results. On the other hand, Model 1, where the impact parameter between the Clouds remains larger than the radius of the LMC's stellar disk, is inconsistent with the observational results in both disk heating and kinematic asymmetry; RMS of the residual PM amplitudes is too small (0.027~\masyr) and asymmetry is not clearly seen in the residual PM field. Similarly, the PM residual for an isolated, unperturbed LMC disk are also inconsistent with the data, indicating that the present-day LMC disk is not in dynamical equilibrium. 

Our comparisons of the observed and simulated LMCs suggest that: 
\begin{enumerate}
    \item A direct collision (i.e., impact parameter $<$ LMC disk size) is needed to simultaneously reproduce the observed level of disk heating and asymmetry in the stellar kinematics. Thus, Model 1 (large impact parameter 20~kpc about 100~Myr ago) can be safely excluded. This further indicates that the tidal field of the Milky Way alone is insufficient to reproduce the asymmetry and level of disk heating observed in the LMC disk; a direct collision with the SMC is required.
    \item For an LMC with a strong direct SMC collision (impact parameter of 2~kpc), the impact must have occurred at least 140~Myr ago in order for the LMC disk to have sufficient time to dynamically cool down and reach a similar dynamical state as that observed in the present-day LMC. Thus, Model 2 (only 100~Myr since the collision) can also be safely excluded. This timing for a direct collision is consistent with independent constraints from the LMC-SMC orbital modeling \citep{Zivick2018} and with the ages of young stars in the Magellanic Bridge \citep{Harris2007, Noel2015}.
    \item Perturbations induced by a direct collision (with any impact parameter $<$10~kpc) will be significantly damped after 250~Myr and become unable to reproduce the present-day LMC's level of disk heating. This sets a stringent upper limit on the timing of a direct collision between the MCs based on dynamics.
    \item Based on our PM residual analysis, we conclude that the most recent encounter with the SMC 
    is direct, with an impact parameter $\lesssim$ 10~kpc and timing within the past 250~Myr. If we adopt the timing constraints from the analytic orbit modeling of \citet{Zivick2018} and the star formation history in the Bridge \citep[e.g.,][]{Harris2007, Noel2015}, our combined results suggest the recent collision occurred  $\sim$140--160~Myr ago with an impact parameter of $\sim$5~kpc, meaning that the recent close encounter also inevitably coincides with the time when the SMC crosses the LMC disk plane.
\end{enumerate}

Given that analytic orbital modeling methods cannot account for the distortions to the dark matter halos of the LMC or SMC, we advocate that the PM residuals presented in this paper are a novel and important constraint that should be used to assess the validity of numerical simulations with a given impact parameter and timing of the LMC-SMC collision.

\acknowledgments
This work presents results from the European Space Agency (ESA) space mission Gaia. Gaia data are being processed by the Gaia Data Processing and Analysis Consortium (DPAC). Funding for the DPAC is provided by national institutions, in particular the institutions participating in the Gaia MultiLateral Agreement (MLA). The Gaia mission website is https://www.cosmos.esa.int/gaia. The Gaia archive website is https://archives.esac.esa.int/gaia. We thank Andr\'es del Pino for useful discussion. G.B. acknowledges support from the NSF under grant AST 1714979. N.K. is supported by NSF CAREER award 1455260.

\software{scipy \citep{jones01}, numpy \citep{vanderwalt11}, matplotlib \citep{hunter07}, astropy \citep{astropy}, emcee \citep{DFM2013}, Project Jupyter (https://jupyter.org), and lmfit \citep{matt_newville_2021_4516651}}

\clearpage

\appendix
\restartappendixnumbering
\section{Appendix}
\begin{figure*}[ht] 
 \centering
      \includegraphics[width=20cm, trim=3.5cm 0cm 0cm 0cm, clip=true]{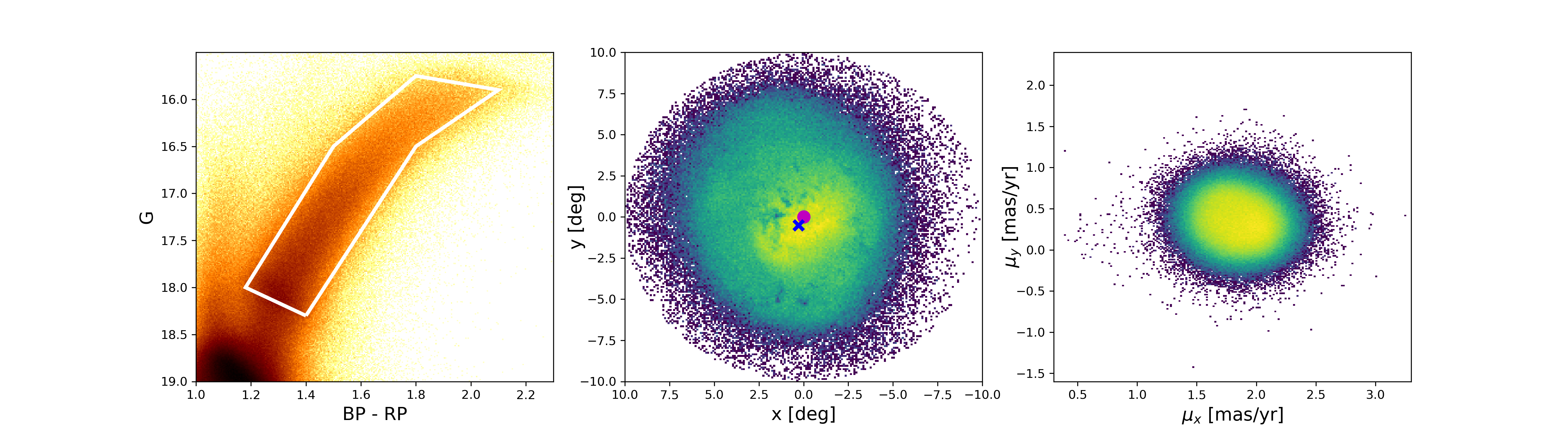}
      \caption{Same as Fig.~\ref{fig:rc_selection}, but for the RGB selection.\label{fig:rgb_selection}}
\end{figure*}

To ensure that our results are indeed independent of tracers among old stellar populations, we also perform the same analyses using the RGB stars, which are brighter than the RC stars, and thus have smaller \textit{Gaia} PM measurement errors and higher completeness \citep{Luri2021}. We select the RGB stars using the same spatial and \textit{Gaia} cuts to the RC stars, but with a different CMD selection. The RGB selection polygon in the CMD is defined as follows: (BP-RP, G) = ((1.4, 18.3), (1.8, 16.5), (2.1, 15.9), (1.8, 15.75), (1.5, 16.5), (1.18, 18.0), (1.4, 18.3)). Figure~\ref{fig:rgb_selection} describes our RGB sample selection. This selection criteria successfully exclude stars with \texttt{ruwe} $>$ 1.23 and \texttt{astrometric\_excess\_noise\_sig} $>$ 2, resulting in a total of 488,795 RGB stars in our final sample. Despite the smaller sample size compared to the RC sample, the fraction of pixels within 6 degree with residual PM S/N $<$ 3 is only $\sim$2\%.

\begin{figure*}[t]
 \centering
      \includegraphics[width=20cm, trim=3.5cm 0cm 0cm 0cm, clip=True]{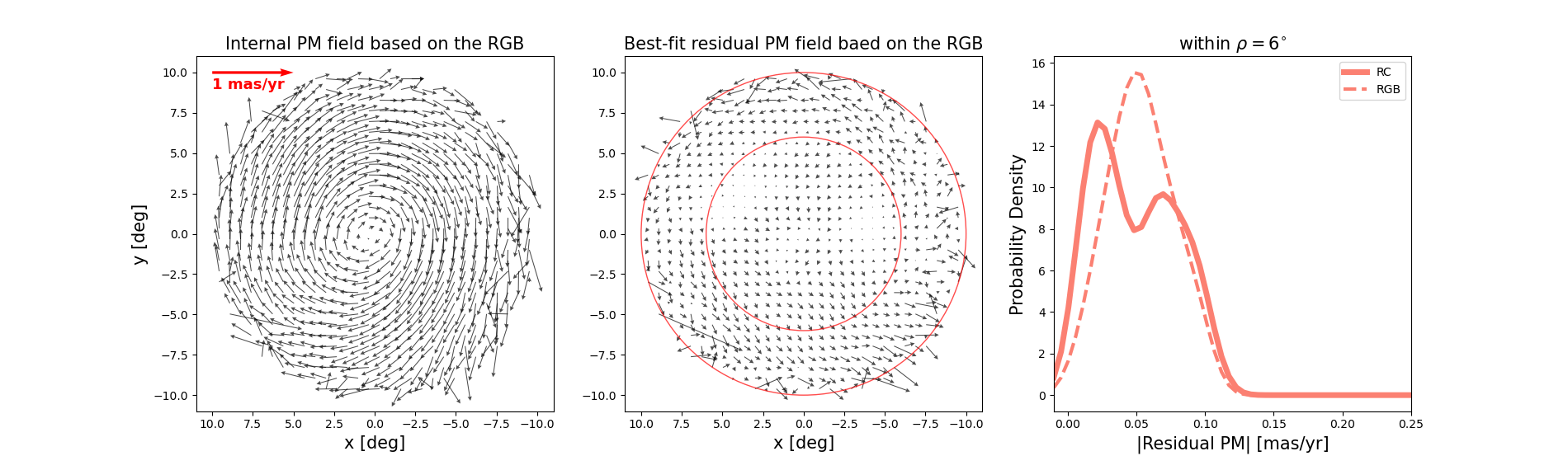}
       \caption{Same as Fig.~\ref{fig:obs_pm}, but for the RGB sample. The first two panels show the internal motion field and residual PM field of the RGB stars. On the third panel, we compare the distribution of the residual PM amplitudes of the RC (solid curve) and RGB (dashed curve) stars within the inner 6\degree. 
      \label{fig:obs_pm_rgb}}
\end{figure*}

In Figure~\ref{fig:obs_pm_rgb}, we present the internal motion and residual PM fields of the RGB sample in the left and middle panels, respectively. The RGB internal motion field also shows a well-organized disk rotation as seen in the RC sample (see Fig.~\ref{fig:obs_pm}).The residual PM field shows the same general behaviors as the RC sample -- larger residuals beyond 6\degree due to the declining rotation curve and asymmetric features. Within the inner 6\degree, one noticeable difference between the RC and RGB samples is that the kinematic asymmetric features appear in different parts of the disk. While the RC residual PM field shows a rough north-south dichotomy, the RGB one is more stochastic with a couple of lower residual spots in the East and West sides of the disk and south east of the bar, leading to a much smoother distribution of the residual PM amplitude than that of the RC sample, i.e., no bimodality. To understand the origin of the spatial discrepancy in the residual PM fields between the RC and RGB populations, a more complex analysis might be required, which is the beyond the scope of this paper. Nevertheless, the RMS of the distribution for the RGB sample is 0.058$\pm$0.002~\masyr, suggesting the same level of the disk heating as we measure with the RC sample. Due to this consistent RMS value, our constraints on the impact timing and parameter remain the same for the RGB sample.

In summary, our main conclusions do not depend on the choice of tracers between the RC and RGB populations. This is consistent with the study by \citet{Luri2021} where they showed that their RC and RGB subsamples from \textit{Gaia} EDR3 share very similar stellar kinematics. 

\bibliography{reference}{}
\bibliographystyle{aasjournal}

\end{document}